\documentclass[useAMS,usenatbib]{mn2e}

\usepackage{graphicx}

\newcommand{\DxDy}

\title[Wave-Field Interaction]{On the Interaction of Internal Gravity Waves with Magnetic Field II. Convective Forcing}
\author[T.M. Rogers and K.B. MacGregor]{T.M. Rogers$^{1}$\thanks{E-mail:tami@lpl.arizona.edu}, K.B. MacGregor$^{2}$\footnotemark[1]\\
$^{1}$Department of Planetary Sciences,University of Arizona, Tucson, AZ 85721\\
$^{2}$High Altitude Observatory, NCAR Boulder, CO 80301}

\begin{document}
\maketitle

\begin{abstract}
We present results from numerical simulations of the interaction of internal gravity waves (IGW) with magnetic fields in the radiative interior of the Sun.  In this second paper, the waves are forced self-consistently by an overlying convection zone and a toroidal magnetic field is imposed in the stably stratified layer just underneath convection zone.  Consistent with the results of previous analytic and simple numerical calculations, we find a strong wave-field interaction, in which waves are reflected in the field region.  The wave-field interaction and wave reflection depend on the field strength as well as adopted values of the diffusivities.  In some cases wave reflection leads to an increased mean flow in the field region.  In addition to reproducing some of the features of our simpler models, we find additional complex behaviours in these more complete and realistic calculations.  First, we find that the spectrum of wave generation, both in magnetized and un-magnetized models, is not generally well described by available analytic models, although some overlap does exist.  Similarly, we find that the dissipation of waves is only partially described by the results of linear theory.  We find that the distortion of the field by waves and convective overshoot leads to rapid decay and entrainment of the magnetic field which subsequently changes the wave-field interaction.  In addition, the field alters the amount of wave energy propagating into the deep radiative interior, at times increasing the wave energy there and at others decreasing  it.  Because of the complexity of the problem and because the durations of these simulations are shorter than the anticipated timescale for dynamical adjustment of the deep solar interior, we are unable to draw a definitive conclusion regarding the efficiency of angular momentum transport in the deep radiative interior by IGW in the presence of a magnetic field.  

\end{abstract}

\section{Introduction}

It has long been known that internal gravity waves (IGW) can transport angular momentum over long distances.  In the Earth's equatorial stratosphere the dissipation of IGW are the primary driving mechanism of the Quasi-Biennial Oscillation (QBO) \citep{ba01} in which the mean zonal flows oscillate with a period of approximately 27-28 months.  The ability of IGW to drive mean flow oscillations and hence, its relevance to the QBO, was demonstrated in the experiment by \cite{pm78}.  

Because of their ability to transport angular momentum over long distances and their anticipated presence in stellar radiative interiors, there is considerable interest in IGW induced transport and mixing processes in stellar interiors .  Nearly 30 years ago \cite{pre81} studied the propagation and dissipation of IGW in radiative interiors, showing that these waves could possibly lead to increased opacity and species mixing.  \cite{gls91} (hereafter referred to as GLS) showed that mixing by IGW could contribute to the enhanced Lithium depletion observe in main sequence F-type stars.  Schatzman has contributed numerous studies on the effects of IGW in stellar interiors.  He has shown that their properties allow a host of transport possibilities, such as the ability to induce diffusion and hence mix species \citep{sch96,mosc00,mosc96}, transport angular momentum \citep{sch93} and even affect opacities and hence, the solar neutrino problem \citep{sch99}.

More recently, internal waves in stellar radiative interiors have received significant attention in connection with claims \citep{kq97,ztm97} that they might be the mechanism responsible for the uniform rotation of the solar interior, as inferred by helioseismic inversion \citep{th96,th03}.  These claims recieved heavy scrutiny \citep{ri98,gm98} because of the known propensity for IGW to force shear flows rather than uniform flows \citep{pm78,ba01}.  More recent incarnations of the IGW theory for the uniform rotation of the solar radiative interior have incorporated an oscillating shear layer, coined the SLO (shear-layer oscillation) as a filter on IGW propagating to the deep interior \citep{ktz99,tkz02,charbonnel05} (we will hereafter refer to \cite{ktz99} as KTZ). However, the validity of these models has also been questioned \citep{rg06,rmg08,den08}, again because the general propensity of IGW to enhance shear layers.  In all of these works the main uncertainty arises from our incomplete understanding of the spectrum and amplitudes of IGW generated.  

While there have been numerous studies on the propagation, dissipation and transport properties of IGW in stellar interiors, very few \citep{sch93,macr10,rmac10} have addressed the issue of magnetism.  The main mechanism thought to generate IGW in the solar radiative interior is convective overshoot at the base of the convection zone.  This is the likely storage region for the strong toroidal magnetic field which gives rise to the sunspot cycle.  Indeed, the sites of IGW generation and magnetic field storage are likely to coincide in many stars.  As pointed out by \cite{sch93} a magnetic field could prevent the propagation of IGW into stellar interiors. For these reasons it is crucial to understand the generation, propagation and dissipation of IGW in the presence of magnetic field.  We began this study with analytical models \citep{macr10} and simple numerical simulations \citep{rmac10}.  Here we extend those studies to more realistic numerical simulations in which the IGW are self-consistently driven by an overlying convection zone.  We then study the interaction of these waves with an imposed toroidal field.  In Section 2 we describe our numerical model, in Section 3 we discuss the generation and propagation of IGW in the presence (3.2) and absence (3.1) of a magnetic field.  In Section 4 we discuss the magnetic spectrum that is generated by IGW impinging on a magnetic field, while in Section 5 we discuss the wave energy reaching the deep radiative interior.  The effect of these waves on the development of the mean flow is discussed in Section 6.  We conclude with a discussion of the wave-field interaction.

\section{Numerical Model}
We solve the coupled set of nonlinear magneto-hydrodynamic (MHD) equations in the anelastic approximation.  The anelastic approximation is valid when the flow velocities and Alfven speed are much smaller than the sound speed.  This approximation effectively filters sound waves, allowing a larger numerical time step.  The equations are solved in a two-dimensional (2D) domain representing an equatorial slice of the Sun that spans the radius range from 0.05R$_{\odot}$ to 0.95R$_{\odot}$.  The details of the equations can be found in \cite{rg05} and \cite{rmac10}, which we will hereafter refer to as P1.  The fundamental difference between this paper and P1, is that here we self-consistently solve for the convection, so that there is no need to artificially force waves on the top boundary.  The radiative region extends from 0.05R$_{\odot}$ to 0.71R$_{\odot}$ and the convection zone extends from 0.71$R_{\odot}$ to 0.90R$_{\odot}$.  For numerical purposes we impose another stable region on top of the convection zone, extending from 0.90$R_{\odot}$ to 0.93R$_{\odot}$.  The reference state thermodynamic variables are taken from the standard solar model by \cite{jcd91}.  

The equations are solved in cylindrical coordinates.  The radial dimension is discretized using a finite difference scheme.  The variables are expanded as a Fourier series in the longitudinal dimension.  The grid-space resolution for all the models is 2048 longitudinal points and 1500 radial points, with 1000 radial levels dedicated to the convection zone, convective-radiative interface and the magnetized region.  Solutions are time-advanced using the Adams-Bashforth method for the nonlinear terms and the Crank-Nicolson method for the linear terms.  The velocity boundary conditions are impermeable and stress-free.  The thermal boundary conditions are isothermal and the magnetic boundary conditions are perfectly insulating. The model is parallelized using Message Passing Interface (MPI).  

As described in P1, the magnetic field is written in terms of a vector potential, resulting in an evolution equation for the vector potential.  In all magnetic models we specify a magnetic field between 0.68R$_{\odot}$ and 0.71R$_{\odot}$, with the following form:

\begin{equation}
A(r)=A_{o}exp^{-(\frac{r}{R_{\odot}} - 0.695)^{2}/0.0002}
\end{equation}
where A(r)$\hat{\bf{z}}$ represents the vector potential, so that the magnetic field (${\bf B}=\nabla\times {\bf A}$) is purely toroidal and is initially confined between 0.68R$_{\odot}$ and 0.71R$_{\odot}$, switching sign at 0.695R$_{\odot}$.  $A_{o}$ is varied substantially, leading to peak field strengths between $200$ and $2\times 10^{5}$G for the various models (see Table 1).\footnote{While the peak field strengths vary substantially, one should keep in mind that, in every model, the field strength varies from 0 to the peak value in the form described in equation 1.}

The control parameters in these models are the magnetic field strength, specified by A$_{o}$, and the thermal ($\kappa$), viscous (${\nu}$) and magnetic ($\eta$) diffusivities.  The magnetic diffusivity is specified to be constant in radius, while both the thermal and viscous diffusivity are functions of radius with the following form: 
\begin{equation}
\kappa=\kappa_{m}\frac{16\sigma {T(r)}^{3}}{3{\rho(r)}^{2}\chi c_{p}}
\end{equation}
\begin{equation}
\nu=\frac{\nu_{m}}{\rho(r)}
\end{equation}
where $\sigma$ is the Stefan-Boltzman constant, T(r) is the reference state temperature, $\rho(r)$ is the reference state density, $\chi$ is the opacity and $c_{p}$ is the specific heat at constant pressure.  Therefore, we vary $\kappa$ and $\nu$ by varying $\kappa_{m}$ and $\nu_{m}$. Because the thermal diffusivity fixes the heat flux through the system and hence, the convective driving, for most models we keep $\kappa_{m}$ (and hence, $\kappa$) fixed.  To investigate the effect of diffusivities and Prandtl numbers, we therefore vary $\nu_{m}$ and $\eta$.  For all magnetic models we run a non-magnetic model for comparison, a list of the models run and their control parameters (diffusivities) are presented in Table 1.  

\begin{table*}
\centering
\begin{minipage}{150mm}
\caption{Model parameters.  $A_{o}$ is the amplitude described in equation 1.  Note that values of $1.66\times 10^{14}$ correspond to peak magnetic field strengths of $\approx 10^{5}G$.  $\kappa$, $\nu$ and $\eta$ are the thermal, viscous and magnetic diffusivities listed in units of $\times 10^{11} cm^{2}s^{-1}$, at a radius of 0.69R$_{\odot}$, equivalent to mid-depth of the imposed magnetic field (where applicable).  Models designated NF* refer to models without a magnetic field, models KV* similarly have no magnetic field but vary $\kappa$.  Models designated BV* vary magnetic field strength keeping diffusivities constant.  Models designated EV* vary $\eta$, and models designated NV* vary $\nu$.}

\begin{tabular}{lllll}

\hline
Model & $A_{o} (10^{14}) (Gauss/cm)$& $\kappa$($10^{11}cm^2/s$) & $\nu(10^{11}cm^2/s$) & $\eta(10^{11}cm^{2}/s$)\\
\hline
NF1  & NA & 64 & 23.6 & NA\\
NF2  & NA & 64 & 18.9 & NA\\
NF3  & NA & 64 & 14.2 & NA\\
NF4  & NA & 64 & 9.5 & NA\\
NF5  & NA & 64 & 4.7 & NA\\
BV1  & 1.660 & 64 & 23.6 & 10.\\
BV2  & 0.166 & 64 & 23.6 & 10.\\
BV3  & 0.0166 & 64 & 23.6 & 10.\\
BV4  & 0.00166 & 64 & 23.6 & 10.\\
EV1 & 1.66 & 64 & 23.6 & 1.\\
EV2 & 1.66 & 64 & 23.6 & 20.\\
EV3 & 0.166 & 64 & 23.6 & 1.\\
EV4 & 0.166 & 64 & 23.6 & 20.\\
NV1 & 1.66 & 64 &18.9& 10.\\
NV2 & 1.66 & 64 &14.2& 10.\\
NV3 & 1.66 & 64 &9.5& 10.\\
NV4 & 1.66 & 64 &4.7& 10.\\
KV1 & NA & 90 & 4.7&NA\\
KV2 & NA & 128 &4.7&NA\\
\hline
\end{tabular}
\end{minipage}
\end{table*}

\section{The Interaction of Internal Gravity Waves and Magnetic Field}

To carry out this study we start with a fairly evolved model from \cite{rg05}, and impose the field described in Equation 1, as seen in Figure 1.  The field is imposed in an already evolved model because the initial time required to get convection started and waves set up is long enough that an initially imposed (two-dimensional) field will have decayed by some amount.  The drawback of this method is that the field is placed in a region which is not quiescent and therefore immediately feels the presence of small scale structure. 
In these simulations it is impossible to disentangle ``wave'' from any other small scale motion.  We therefore use the term ``wave'' to mean the following: if a variable is Fourier decomposed in longitude (as we do in these simulations), ``wave'' is any horizontal wavenumber, k, not equal to zero.  Mean flow therefore, refers to horizontal wavenumber, k, equal to zero.  

\subsection{Waves in the absence of a Magnetic Field}
One of the central questions with regard to IGW transport and mixing is the spectrum and amplitude of waves generated at the convective-radiative interface.  While there is still some debate, previous numerical calculations which include the most relevant physics, but which are unfortunately limited to 2D \citep{rg05}, have predicted a broad spectrum of IGW generated at the base of the convection zone.  Indeed, the spectrum is very similar to that of convection.  Deeper in the radiation zone the spectrum is dominated by high-frequency, low-wavenumber waves.  Standing modes appear to be short-lived.  

Figure 2 shows the spectrum of horizontal wavenumber, k, versus frequency\footnote{Note that for most of the spectra shown the frequency resolution is 0.97$\mu$Hz.  However in 3.2 we study the effect of decaying field and in that case the frequency resolution is reduced.} for Model NF5 (no magnetic field) at three radii, 0.71$R_{\odot}$, 0.70$R_{\odot}$ and 0.69$R_{\odot}$.  The color tables represent kinetic energy density.  One can see that motion at the base of the convection zone has its highest amplitudes at lower wavenumbers and frequencies; there is little decrease in amplitude until wavenumber $\approx$ 200 and frequencies of $\approx 100-150\mu$Hz.  

The frequency and wavenumber dependence of the spectrum of motions at the base of the convection zone can be seen more clearly in Figure 3, which shows the wave energy density ($\rho(v_{r}^{2}+v_{\phi}^{2})$) as a function of frequency (a) and wavenumber (b) at a radius of 0.70R$_{\odot}$, for a variety of models.  Figure 3a shows a strong dependence of wave generation on frequency, with the lowest frequencies generated with the highest amplitudes and the highest frequencies nearly three orders of magnitude lower in amplitude, over a frequency range of 100$\mu$Hz.  In Figure 4 we show fits to these curves for a variety of horizontal wavenumbers for model NF5; we find that the frequency dependence of the velocity amplitude can best be described as decaying like $\omega^{-2.7}$, although we find substantial variance, both with frequency and horizontal wavenumber, with exponents ranging from $\approx $-2.2 to -3.2.  If we compare this spectrum to that in the convection zone, we find very little difference (other than amplitude, see Figure 4 dashed line).\footnote{The similarity of the spectrum to that in the convection zone is not surprising since continuity of pressure at the interface implies continuity in velocity.} There is some evidence for two slopes in the spectrum, with high frequencies better fit by a slightly steeper power law of $\omega^{-3.2}$ and lower frequencies fit by $\omega^{-2.2}$ which may indicate different generation mechanisms.  However, there is enough variance of the frequency dependence with horizontal wavenumber that it is difficult to draw any firm conclusions about generation mechanism.  Since the energy density varies as approximately $\omega^{-2.7}$ and the group velocity varies as approximately $\omega$, we estimate that the wave energy flux varies as approximately $\omega^{-1.7}$.  

Clearly, the amplitudes of the motions driven at the base of the convection zone depend on frequency.  However, in Figure 3b we see that the generation of wave energy shows a very mild dependence of excitation on horizontal wavenumber with a maximum variation of one to two orders of magnitude (and generally less) over a horizontal wavenumber range of 100.  At low wavenumbers, the amplitudes increase with horizontal wavenumber, similar to that predicted analytically in GLS and KTZ.  However, from horizontal wavenumber 20-50, the amplitudes are relatively flat, and beyond that the amplitudes decrease.  Motions in the convection zone show a similar, relatively flat dependence of energy on wavenumber, at least out to horizontal wavenumber 100-200.

There has been significant discussion about the generation of IGW at the base of the solar convection zone.  In both the analytic works by GLS and KTZ the spectrum is calculated by considering the stress imposed at the convective-radiative interface by convective eddies with smaller scales described by a Kolmogorov cascade.  Another likely, and probably dominant, mechanism for generating IGW is that of overshooting plumes for which there is no analytic prediction of the spectra.  In these simulations, we have both stresses at the interface due to convective eddies, as well as convective overshoot by plumes.  The spectrum generated in these simulations with the energy density varying as $\omega^{-2.7}$ and non-monotonic variation with wavenumber, is unlike either GLS or KTZ, which both predict energy density varying as $\omega^{-2}k$.

There could be many sources for the discrepancy between the spectrum produced by these simulations and those of analytic models.  For example, both KTZ and GLS assume a Kolmogorov cascade which describes isotropic, inhomogeneous turbulence, which is not likely representative of the base of the solar convection zone.  The discrepancy could be due to the two-dimensional nature of these simulations compared to the one dimensional description of those analytic models.  Nonlinear interactions between waves included here but neglected in those models could give rise to different correlation scales and times and therefore, a different spectrum.  However, we think the most likely source for the discrepancy is the inclusion in these models of overshooting plumes.  Unfortunately, in this model we can not discern (and possibly it is not useful to do so) which mechanism (stresses and deformation of the interface versus overshooting plumes) is responsible for the spectrum observed.  It is plausible that the limited extent of a plume incursion could give rise to a broader wavenumber spectrum and that the short timescale of such an event leads to higher frequencies.  However, our spectrum is not completely different than that obtained from analytic models and therefore, it appears that the most likely scenario is that both mechanisms are relevant.

In addressing the spectrum of waves generated we can look to experimental results in limited circumstances.  Experiments of IGW generated by an overshooting plume \citep{ansuth10}, have predicted that the frequencies generated correspond to a narrow range (0.6-0.8) of the ${\it constant}$ buoyancy frequency, N.  This range is similarly found with sinusoidal disturbances dragged along the boundary \cite{agsuth06}.  While these results are interesting and likely relevant, it is hard to make the comparison with the models presented here, or the Sun for that matter, as the buoyancy frequency varies substantially with radius.\footnote{If one were to take this seriously, the range of 0.6-0.8 N, with N varying between zero and a few hundred $\mu$Hz, could give a broad frequency range as seen here.}  

In figure 2 we can also see rapid dissipation of the waves over the small radial range shown, we can particularly see low frequencies and higher wavenumbers dropping in amplitude with radius, consistent with simple radiative dissipation.\footnote{Clearly, the dissipation here is stronger than is likely in stellar interiors because of our necessarily large diffusion coefficients}.  We can investigate the dissipation of waves as a function of radius more readily by looking at particular frequency and wavenumber combinations as a function of radius.  In Figure 5a the dissipation of waves with radius is shown for three different models with color representing either varying thermal diffusivities (top four panels) or varying viscous diffusivities (bottom four panels) and linetypes indicating different frequencies, all with the same wavenumber (k=10,a,e,c,g) or different wavenumbers all with the same frequency (10$\mu$Hz,b,f,d,h).  The left hand panels (a-d) are not adjusted for the variation in generation amplitude, discussed above, while the right hand panels (e-h) are adjusted.  We can see two things immediately: (1) the wave dissipation is fairly independent of $\nu$ and $\kappa$ and (2) the wave amplitudes in the deep radiative interior are fairly ${\it independent}$ ${\it of}$ ${\it frequency}$.  We are careful here to distinguish that the wave amplitudes in the deep interior are independent of frequency because the wave ${\it dissipation}$ is dependent on frequency.  This apparent contradiction arises because wave generation is strongly dependent on frequency, so that lower frequencies are generated at higher amplitude, but are simultaneously dissipated much more strongly, leading to very little difference in wave amplitude in the deep radiative interior across a large range of frequencies\footnote{Of course this is not strictly true.  As can be seen in the figure significantly higher frequency waves have higher amplitudes in the deep interior}.  Indeed, if we correct for the generation amplitude we can see a clear frequency dependence in Figure 5 (e,g), with higher frequencies damped less than lower frequencies, as expected from simple radiative dissipation.

We can calculate the dissipation length by measuring that radius at which the wave amplitude decays by a factor of ${\it e}$, this is shown in Figure 6.  In that figure the damping length is shown as a function of frequency and various linetypes depict different horizontal wavenumbers.  We immediately see that the linear description of waves is ${\it qualitatively}$ recovered; higher frequencies have longer damping lengths.  However, we do not find the severe $\omega^{4}$ dependence predicted, but instead find something close to $\omega^{3}$ ($\omega^{2.7}$ is shown in the figure).  This is qualitatively understood by considering that the damping length is proportional to the group velocity divided by some damping rate, as outlined in \cite{ktz99}.  In the case of the Sun the ratio $\omega/N$ is quite small and therefore the wave path is ${\it mostly}$ horizontal.  Therefore, one should consider the ${\it horizontal}$ group velocity $\partial{\omega}/\partial{k_{h}}$, given approximately by $N/m$, where m is the vertical wavenumber.  On the other hand, dissipation is likely more dominant in the vertical direction where the gradients are largest.  Therefore, the dissipation rate is given by $\kappa k_{v}^{2}$.  This leads to a damping length which is given by $\approx \omega^{3}/(N^{2}k_h^{3}\kappa)$.  We should note here that our calculated damping lengths only recover the quantitative behaviour in damping length for a limited range of frequency and wavenumber combinations.  At very low frequencies and high wavenumbers the vertical wavelength is relatively small and likely not well resolved throughout the domain (in the deep radiative interior we resolve vertical wavelengths of approximately $10^{8}cm$, which if linear theory holds, would require $\omega/k$ to be greater than $\approx 10^{-8}$ for N $\approx 100\mu$Hz).  On the other hand high frequencies and low wavenumbers have relatively large vertical wavelengths and the WKB approximation is not well justified.\footnote{For frequencies of 100$\mu$Hz and horizontal wavenumber 5, the ratio of $\lambda_r(\partial{N^{2}}/\partial{r})/N^{2}$ is larger than 0.01 for most of the radiative interior}.  

Another potential problem with calculating the damping length for a variety of frequencies and wavenumbers is that for many frequency/wavenumber combinations the motion does not appear ``wave-like'' until fairly deep into the radiative interior.  This is also apparent as two distinct slopes for many frequency/wavenumber combinations seen in Figure 5.  Nearer the convection zone the slope is steeper than deeper into the radiative interior.  This is because close to the convection zone the waves are generated with rather high amplitudes and a simple estimate of the linearity of the waves given by $c_{p}/u \gg $ 1 does not hold and linear theory is likely not valid.  In this region mode-mode coupling is efficient and wave energy can be transferred to shorter scales where it is dissipated rapidly.  Deeper into the radiative interior, wave amplitudes are diminshed and one can clearly discern wave like structure.  Here, wave-wave coupling in scale is less efficient, leading to dissipation roughly in agreement with simple predictions.  

While the viscous diffusivity plays little role in the dissipation of waves with radius within the radiation zone, it does affect the amplitudes of the waves generated.  Lower values of the viscous diffusivity lead to less dissipation of the convective motions and therefore, slightly larger convective velocities.  This in turn, leads to slightly higher amplitude motions below the convection zone at all frequencies and wavenumbers, as can be seen in Figure 7 a,b,c.  In particular, lower values of the viscous diffusion lead to higher amplitudes at higher frequencies.  This is likely due to more dominant nonlinear terms (accompanying the larger Reynolds number) which lead to more efficient wave-wave coupling. This is seen in Figure 7d where, for a given frequency, lower values of the viscous diffusivity show shallower slopes in horizontal wavenumber, indicating better transfer between various scales.  Moreover, nonlinear interactions between waves could change the correlation times of those waves explaining the change in the frequency dependence of the spectrum.
  
\subsection{Waves in the presence of a Magnetic Field}

In \cite{macr10} it was shown that the strength of the magnetic field-IGW interaction depends on the difference between the Alfven frequency and the IGW frequency.  If one considers a vertically propagating wave travelling from region 1 with field strength described by an Alfven velocity, $u_{A1}$, to an underlying region 2 with field strength described by $u_{A2}$, then the reflection coefficient is given by (equation (27) that paper):
\begin{equation}
R=\frac{1-q}{1+q}
\end{equation}
with 
\begin{equation}
q=\frac{m_{2}}{m_{1}}(\frac{\omega^{2}-k^{2}u_{A2}^{2}}{\omega^{2}-k^{2}u_{A1}^{2}})
\end{equation}
Here m$_{2}$ and m$_{1}$ represent the vertical wavenumbers in the two regions, $\omega$ the IGW frequency, and k is the horizontal wavenumber.  The reflection coefficient thus depends on the ratio of the field strengths in region 1 and 2 (or, in the case presented here, on the derivative of the field strength with radius) and the IGW wave frequency.  It was shown in \cite{macr10} that for low field strengths the reflection coefficient is small, but non-zero for a wide range of IGW frequencies.  On the other hand, for high field strengths the reflection coefficient is large (one) but for a narrower range of IGW frequencies.  This is generally true for both steep and shallow field strength gradients.  Since these results imply that reflection could occur for a range of field strengths and IGW frequencies, we first turn our attention to how the IGW spectrum is altered in the presence of a magnetic field.

When looking at the absolute spectrum of waves at some radius below the convection zone it is difficult to discern the small differences in amplitudes between different models.  Therefore, in the following we will often show the ${\it ratio}$ of spectra for different models.  We show the ratio of the ${\it kinetic}$ energy spectrum of waves for Model BV1 (strong magnetic field) to that of NF1 (no field), in Figure 8a and the ratio of the kinetic energy spectrum of waves for Model BV2 (weak magnetic field) to that of Model NF1 (no field) in Figure 8b at a radius of 0.69R$_{\odot}$.  These two models have all other parameters equivalent.  In this figure regions which are blue indicate regions in frequency and horizontal wavenumber (f,k) space where the ratio is close to zero, meaning there is little or no energy in that region for the magnetic case, relative to the non-magnetic case.  Regions which are red represent ratios of $\geq$ 1, implying that there is as much or more energy in that region of (f,k) space in the magnetic case as the non-magnetic case.  What is immediately obvious in 8a are the dark ridges.  In Figure 8b one can see dark ridges, but these features are both less severe and spread over a larger range in frequency and wavenumber.  This is similar to what is seen in the analytic results of \cite{macr10}.  Namely, strong magnetic fields lead to strong reflection in a limited region of (f,k) space, while weak fields lead to weaker reflection in a larger region of (f,k) space.  

As described in Section 2 we impose a magnetic field which is purely toroidal. In the absence of a dynamo, the field cannot regenerate itself, and we would expect this toroidal field to decay in a diffusion time given by $\tau_{diff}=D^{2}/\eta$, where D is the depth of the magnetic layer.  In our simulations the field is distorted by overshooting motions and IGW and therefore, the timescale over which diffusion occurs is much shorter.  We show this effect in Figure 9, which displays the square of the mean toroidal magnetic field as a function of time in our simulations (solid lines) for Models EV1 (black), BV1 (blue) and EV2 (red) and that expected for pure diffusion, with the specified diffusion rate (dashed lines).  Clearly, our field is decaying much more rapidly than expected for pure diffusion.  In fact, if one were to calculate the ``effective magnetic diffusivity'', $\eta_{eff}$, for each of these models one finds that Model EV1 has $\eta_{eff}\approx 10\eta_{ev1}$, Model BV1 has $\eta_{eff}\approx 3\eta_{ev1}$ and Model EV2 has $\eta_{eff}\approx 2\eta_{ev2}$.  The dependence of the enhancement on $\eta$ is reflective of the fact that higher initial values of $\eta$ represent fields which are not as well tied to flows as their lower $\eta$ counterparts and the field therefore, responds less to flow imposed distortion.  

Because filtering of the spectrum is a strong function of field strength, and because the field strength is a function of time, we need to bear in mind that the wave filtering described above, could also be a strong function of time.  We demonstrate this in Figure 10, where we show wave filtering as a function of time.  In Figure 10 a,d,g,j we show the ratio of kinetic energies for the magnetic model compared to the non-magnetic model over a timescale of $\approx 10^{6}$s and therefore, with a frequency resolution of $\approx 1\mu$Hz, in (b,e,h,k) the timescale is $5 \times 10^{5}$s and in (c,f,i,l) the timescale is $2.5 \times 10^{5}s$, with the corresponding decrease in frequency resolution.  Various rows show the ratio for different models (as outlined in the figure and figure caption).  It is immediately clear that the decaying field has a profound effect on the filtering of waves.  In all cases, when the field is relatively in tact (early in the simulation) wave filtering is severe.  For most models, filtering is reduced as the field decays.  The changing properties of wave filtering are likely due to two effects: (1) the field amplitude decays and therefore the reflection coefficient described in equation 5 above decreases and (2) the wave-field interaction becomes less severe as the field is modified by the overlying convection.  

The field evolution for Model NV4 is shown in Figure 11.  The initial toroidal field does not remain stationary for long.  Waves and convective overshoot impinge on the field causing smaller scale structure.  Eventually the field is directly impinged by overshooting plumes which entrain the field into the convection zone.  Once there, the dynamics is best described as magnetoconvection.  Field in the convection zone is subsequently pumped back down and this very non-linear process continues, complicating the simple field-IGW interaction picture.  Of course, once the field is distorted, the field gradient and thus the reflection coefficient become spatially and temporally dependent.  A spatially dependent reflection coefficient could then allow transmission at one longitude while reflecting a wave at another.  

The above discussion describes models in which the magnetic Prandtl number, Pr$_{m}$ (=$\nu/\eta$), is greater than one.  In the solar tachocline Pr$_{m}$ is less than one.  To study the dependence of wave filtering on Pr$_{m}$ we keep $\eta$ fixed and lower $\nu$.  The thermal and viscous diffusivities change substantially over the computational domain, as can be seen from their radial forms in equations 2 and 3, so when we refer to Prandtl (magnetic and non-magnetic) numbers less than one we are referring to the region from 0.69R$_{\odot}$ and 0.71R$_{\odot}$.  All models have Prandtl numbers less than one, and we ran one model for which the magnetic Prandtl number is less than one.  If we compare Figure 10 a,b,c with Figure 10 j,k,l we can see that as Pr$_{m}$ is lowered below one, more wave energy is filtered in the field region.  We can also see that Model BV1, shown in Figure 10 a,b,c is more susceptible to the effects of field diffusion on wave filtering when compared to 10 j,k,l, ${\it even}$ ${\it though}$ ${\it the}$ ${\it magnetic}$ ${\it diffusivity}$ ${\it is}$ ${\it equivalent}$ in the two models, while the viscous diffusivity is substantially lower in NV4.  This indicates the coupled nature of field and flow and their dependencies on ${\it all}$ diffusivities.

The decrease in kinetic energy in particular regions of (f,k) space indicates that incoming wave energy is reflected and ducted between field and no-field regions or between neighboring field regions (as implied in \cite{macr10}).  In addition, it is likely that incoming IGW undergo mode conversion to Alfven waves; indeed, once in the field region gas motion is of mixed gravity-Alfven type.  We now turn our attention to the magnetic spectrum that is generated by the impinging IGW generated by convective overshoot.

\section{Magnetic Spectrum}

Although it is possible that IGW are mode converted to Alfven waves, it is virtually impossible to distinguish that behaviour in these nonlinear calculations.  Therefore, in the following we will discuss the spectrum of magnetic fluctuations that develop from impinging IGW and how that magnetic energy evolves in time, without specific regard to the exact manner in which IGW energy is converted to magnetic energy.  

Alfven waves are described by the simple dispersion relation $\omega_{A}=v_{A}k$, where $v_{A}$ is defined $v_{A}=B_{o}/\sqrt{4\pi\rho}$.  Because the initial field we impose, $B_{o}$, varies substantially with radius, a range of Alfven speeds are possible.  Furthermore, the convective motions drive a wide range of wavenumbers (see Figure 2), so one could expect a very broad range of Alfven frequencies to be generated.  Indeed, if Alfven waves were directly excited by IGW with the corresponding wavenumber, one would expect Alfven frequencies in the range $0 < \omega_{A} < B_{o}k/\sqrt{4\pi\rho}$.  This would lead to frequencies ranging from zero to $\approx 14 \mu$Hz for Model BV3 and from zero to $ \approx 1400 \mu$Hz for Model BV1 (with the caveat of field decay discussed above).  However, the interaction which converts IGW to magnetic energy is nonlinear and therefore depends on a variety of factors and direct excitation does not occur.

As expected from the basic relationship above, we find that models with lower initially imposed field strengths generally produce waves with lower frequencies and amplitudes, as seen in Figure 12.  We see that large(r) amplitude magnetic energy is concentrated at progressivley lower frequencies and wavenumbers for lower field strengths.  Also noteworthy is broadband magnetic energy seen at all frequencies, due to nonlinear interactions.  The models in Figure 12 all have $Pr_{m}$ larger than one.  Lowering $Pr_{m}$ below one also affects both the kinetic energy spectrum and the magnetic energy spectrum, as seen in Figure 13.  Lower magnetic Prandtl number models show a high frequency branch not seen in higher Pr$_{m}$ cases.  When inspecting the ratio of the kinetic spectrum shown in Figure 13a we see that there is a large region of (f,k) space for which the magnetic model does not support (kinetic) motion.  This is the same region for which there appears to be decreased magnetic energy.  This could indicate mode conversion, whereby when an incoming IGW is reflected, the field responds to the reflection at a different wavenumber and frequency.  However, this behaviour is not completely understood.

When wave motion is reflected at a given radius in the magnetized region, the now upward propagating wave is likely to be re-reflected at the top of the magnetized region, when propagating from magnetized to unmagnetized layers, and so on.  The energy in this ducted region could be dissipated, transferred to magnetic energy (as dicussed in this section), or transferred to mean flow as shown in \cite{rmac10}.  Whatever the case, the energy propagating into the deeper radiative interior is affected.

\section{Wave Transmission}

In the above discussion we showed that a toroidal magnetic field below the convection zone could filter the waves propagating in that region.  In the above we investigated how this filtering affects the wave spectra, looking in detail at what frequencies and wavenumbers are filtered, how much they are filtered, and how that filtering depends on field strength and magnetic Prandtl number.  Now we turn to the problem of how that filtering affects the amount of wave energy reaching the deeper radiative interior.  This is a critical consideration because in discussions of the ability of IGW to transport angular momentum or mix species, two main questions arise: (1) the spectrum of waves generated and (2) the amplitudes of waves generated, i.e. how much wave energy is available.   In the section 3.1 above and in \cite{rg05} we addressed (1) and (2) for non-magnetized models, in section 3.2 we addressed how a magnetic field alters (1) and here we will address (2) in the magnetized models.  
 
Figure 14 shows the ${\it integrated}$ kinetic energy in waves from the bottom of the computational domain to 0.67R$_{\odot}$, just beneath the position of the initially imposed magnetic field.  What is shown in the figure is the ratio of integrated kinetic energy in waves in the magnetic case to the integrated kinetic energy in waves for the non-magnetic case.  A value of one would indicate that the magnetic models produce as much integrated wave energy in the deep interior as a non-magnetic case, while a value less (greater) than one indicates that the magnetic models produce ${\it less}$ (${\it more}$) energy in that region.  Initially (up to about $2\times10^{5}s$) the ratio for all models oscillates but stays very close to one.  However, after this initial period (which is approximately the time it would take many of the IGW to propagate through the magnetic layer) there is substantial variation with ratios reaching as low as 0.2 and as high as 4.  

Given the filtering seen in the above sections it is not surprising that the wave energy below the field would be reduced from the non-field case.  However, the fact that there is a substantial period during which the ratio is larger than one is unexpected.  This can be seen more clearly in Figure 15, which shows the kinetic energy in waves as a function of radius and time for the magnetic case (top) and the non-magnetic case (bottom).  In this figure one can clearly see the hallmarks of IGW propagation with upward propagation of phase and downward propagation of energy.  The top and bottom figures look virtually identical for the first $2\times10^{5}s$ (which corresponds to the initial oscillatory behaviour seen in Figure 14), but then one can see a clear change in the field case with subsequent higher amplitude disturbances propagating inward in the magnetic model.  
One interpretation of this behaviour is tht the field, by virtue of reflection (and subsequent reflection at the base of the convection zone) traps wave energy in the field regio, causing the integrated energy in the deep radiative interior to drop below one.  Once the field is sufficiently decayed or distorted the once trapped energy is allowed to propagate into the deeper radiative interior, causing the integrated wave energy in the deep radiative interior to increase again, as seen in Figure 14.  This process could be very time dependent given the spatial and temporal dependene on the field in the region.  This interpretation is somewhat supported by figure 14b, which shows the integrated wave energy ratio for model NV4/NF5 beneath the initially imposed field (as seen in figure 14a, solid line) as well as the ratio of the sum of the integrated wave kinetic and magnetic energy ratio ${\it in}$ ${\it the}$ ${\it field}$ ${\it region}$.  If the above interpretation is correct one would expect that the energy in the field region would be maximum when the energy in the deeper interior is minimum.  This is true during much of the time shown in Figure 14b\footnote{Although, it is clearly not true all of the time.}, making this a plausible interpretation and one which is consistent with the theories laid out here and in \citep{macr10,rmac10}.

In Figure 15 one also notices a great deal of complexity.  Even in the deep radiative interior, where the wave amplitudes have been severely attenuated, there is still considerable interaction leading to significant time and position variation of amplitudes and phase and group speeds.  

\section{The effect of the Magnetic Field on the Mean Flow}
One of our main concerns when undertaking this study was how an imposed field might affect the ability of IGW to transport angular momentum in to the deep radiative interior.  In Figure 16 we show the ratio of the kinetic energy in the mean flow for the magnetized model compared to the non-magnetic model, in the field region (top) and in the deeper radiative interior (bottom), for various models.  One can see that there is very little change in the mean flow in the deep radiative interior.  This could be physical, in the sense that an imposed magnetic field does not affect the correlations which contribute to the mean flow.  However, it is possible that this is a numerical limitation.  The mean flow in the deep radiative interior generally takes a substantial amount of time to grow.  We are limited in how long we can run these models because of the field decay discussed in section 3.2; if we run these models much longer they would eventually revert to a non-magnetic case and the comparison would be moot.  So, while this is an important question to address, we are unable to do so definitively with these models.

In the field region where the timescale for mean flow growth is likely to be much shorter (because the amplitudes of the waves are larger and because the region is shallower), we see that in most models the mean flow changes little from the unmagnetized model.  However, in a few models we see that the mean flow grows and in our (arguably) most realistic model (lowest Pr$_{m}$, largest field strength) we see that the mean flow grows substantially.  This is similar to what was seen in \citep{rmac10}\footnote{Although in that case it occurred even for Pr$_{m}$ =1.}.  This is due to the fact that more wave energy is confined to the region, either by the ducting proposed in \cite{macr10} or by secondary excitation of IGW by field shown in Figure 15.  More wave energy at a variety of wavenumbers and frequencies is likely to lead to more wave-wave interaction which could contribute to the mean flow. 

\section{Discussion}

In this paper we have continued our investigation of the interaction of IGW with magnetic fields.  This paper addresses the complex problem of a wave spectrum self-consistently driven by an overlying convection zone interacting with an evolving magnetic field, similar what is expected in the solar interior.  As expected from our previous work we find that magnetic fields reflect IGW in a way which depends on the magnetic field strength.  This wave-field interaction excites magnetic fluctuations which also depend on field strength in a way which is consistent with the simple Alfven relation. 

Naively, the reflection of waves by a magnetic field would imply that less wave energy would make it to the radiative interior below the magnetic field.  However, we find that this is not always the case.  The wave energy below the field varies substantially in time and can be both larger and smaller than than in a non-magnetic case.  

Although one of our main motivations for this work was to understand how a magnetic field might alter the angular momentum transport in the deep radiative interior, we are cautious about conclusions drawn from these simulations for the mean flow in the deep radiative interior.  This hesitation is because the timescale for the mean flow to evolve in the deep radiative interior is likely to be longer than the duration of these simulations.  In the region of the field we see that, generally, the mean flow is very similar in the magnetized and un-magnetized models.  However, in cases with lower values of $\nu\eta$ we see an enhanced mean flow, due to increased wave energy (by reflection or secondary generation) in the region.

Although many of our results agree with the analytic theory in \cite{macr10} and the simple models in \cite{rmac10}, we find many complications in these models.  First, the field evolves substantially in time so that the simple picture of a stationary field with waves impinging on it is not an accurate description of the dynamics.  The field evolution is dominated by the distortion of the field into smaller scales and the entrainment of field into the convection zone, and therefore is not diffusion dominated.  This evolution affects wave filtering and, through its affect on the convection, wave generation.  These complications (and more) are very likely to exist in the solar interior.  The position of the toroidal field thought to give rise to the sunspots is in a region accessible to convective overshoot in order to allow for magnetic pumping.  Therefore, distortion and entrainment of the field is likely.  Additionally, the gas motions in that region are likely not linear waves, but some combination of nonlinearly interacting waves and convective overshoot as we noted in section 3.1.  Second, even in the deep radiative interior, where wave amplitudes are low there appears to be substantial wave-wave interaction which leads to dissipation rates not well described by linear theory.  Finally, the secondary effect of field generated IGW (or, alternatively, time delayed and amplified waves) make estimates of wave energy in the deep radiative interior highly uncertain, given the unknown field strengths in the region and their evolution in time.  Of course, all of this could be complicated even more by the presence of a primordial poloidal field, which will be the subject of a future work.

\bibliographystyle{mn2e}
\bibliography{igwcl}

\begin{thebibliography}{}

\bibitem[\protect\citeauthoryear{Aguilar \& Sutherland}{Aguilar \&
  Sutherland}{2006}]{agsuth06}
Aguilar D.,  Sutherland B.~R.,  2006, Physics of Fluids, 18

\bibitem[\protect\citeauthoryear{Ansong \& Sutherland}{Ansong \&
  Sutherland}{2010}]{ansuth10}
Ansong J.~K.,  Sutherland B.~R.,  2010, J.~Fluid Mech., 648, 405

\bibitem[\protect\citeauthoryear{Baldwin, Gray, Dunkerton, Hamilton, Hayes,
  Randel, Holton, Alexander, Hirota, Horinouchi, Jones, Kinnersley, Marquardt,
  Sato \& Takahashi}{Baldwin et~al.}{2001}]{ba01}
Baldwin M.,  Gray L.,  Dunkerton T.,  Hamilton K.,  Hayes P.,  Randel W.,
  Holton J.,  Alexander M.,  Hirota I.,  Horinouchi T.,  Jones D.,  Kinnersley
  J.,  Marquardt C.,  Sato K.,    Takahashi M.,  2001, Reviews of Geophysics,
  39, 179

\bibitem[\protect\citeauthoryear{Charbonnel \& Talon}{Charbonnel \&
  Talon}{2005}]{charbonnel05}
Charbonnel C.,  Talon S.,  2005, Science, 309, 2189

\bibitem[\protect\citeauthoryear{Denissenkov, Pinsonneault \&
  MacGregor}{Denissenkov et~al.}{2008}]{den08}
Denissenkov P.,  Pinsonneault M.,    MacGregor K.,  2008, Astrophysical
  Journal, 684, 757

\bibitem[\protect\citeauthoryear{Garcia-Lopez \& Spruit}{Garcia-Lopez \&
  Spruit}{1991}]{gls91}
Garcia-Lopez R.,  Spruit H.,  1991, ApJ, 377, 268

\bibitem[\protect\citeauthoryear{Gough \& McIntyre}{Gough \&
  McIntyre}{1998}]{gm98}
Gough D.,  McIntyre M.,  1998, Nature, 394, 755

\bibitem[\protect\citeauthoryear{J.Christensen-Dalsgaard, Gough \&
  Thompson}{J.Christensen-Dalsgaard et~al.}{1991}]{jcd91}
J.Christensen-Dalsgaard Gough D.,    Thompson M.,  1991, ApJ, 378, 413

\bibitem[\protect\citeauthoryear{Kumar \& Quataert}{Kumar \&
  Quataert}{1997}]{kq97}
Kumar P.,  Quataert E.,  1997, ApJL, 475, 133

\bibitem[\protect\citeauthoryear{Kumar, Talon \& Zahn}{Kumar
  et~al.}{1999}]{ktz99}
Kumar P.,  Talon S.,    Zahn J.,  1999, ApJ, 520, 859

\bibitem[\protect\citeauthoryear{MacGregor \& Rogers}{MacGregor \&
  Rogers}{2010}]{macr10}
MacGregor K.,  Rogers T.,  2010, Solar Physics, submitted

\bibitem[\protect\citeauthoryear{Montalban \& Schatzman}{Montalban \&
  Schatzman}{1996}]{mosc96}
Montalban J.,  Schatzman E.,  1996, Astronomy and Astrophysics, 305, 513

\bibitem[\protect\citeauthoryear{Montalban \& Schatzman}{Montalban \&
  Schatzman}{2000}]{mosc00}
Montalban J.,  Schatzman E.,  2000, Astronomy and Astrophysics, 354, 943

\bibitem[\protect\citeauthoryear{Plumb \& McEwan}{Plumb \& McEwan}{1978}]{pm78}
Plumb R.,  McEwan A.,  1978, J. Atmos. Science, 35, 1827

\bibitem[\protect\citeauthoryear{Press}{Press}{1981}]{pre81}
Press W.,  1981, ApJ, 245, 286

\bibitem[\protect\citeauthoryear{Ringot}{Ringot}{1998}]{ri98}
Ringot O.,  1998, Astronomy and Astrophysics, 335, L89

\bibitem[\protect\citeauthoryear{Rogers \& Glatzmaier}{Rogers \&
  Glatzmaier}{2005}]{rg05}
Rogers T.,  Glatzmaier G.,  2005, MNRAS, 364, 1135

\bibitem[\protect\citeauthoryear{Rogers \& Glatzmaier}{Rogers \&
  Glatzmaier}{2006}]{rg06}
Rogers T.,  Glatzmaier G.,  2006, ApJ, 653, 756

\bibitem[\protect\citeauthoryear{Rogers \& MacGregor}{Rogers \&
  MacGregor}{2010}]{rmac10}
Rogers T.,  MacGregor K.,  2010, Monthly Notices of the Royal Astronomical
  Society, 401, 191

\bibitem[\protect\citeauthoryear{Rogers, MacGregor \& Glatzmaier}{Rogers
  et~al.}{2008}]{rmg08}
Rogers T.,  MacGregor K.,    Glatzmaier G.,  2008, MNRAS, 387, 616

\bibitem[\protect\citeauthoryear{Schatzman}{Schatzman}{1993}]{sch93}
Schatzman E.,  1993, Astronomy and Astrophysics, 279, 431

\bibitem[\protect\citeauthoryear{Schatzman}{Schatzman}{1996}]{sch96}
Schatzman E.,  1996, Journal of Fluid Mechanics, 322, 355

\bibitem[\protect\citeauthoryear{Schatzman}{Schatzman}{1999}]{sch99}
Schatzman E.,  1999, Phys. Rep. Vol. 311, 3, 143

\bibitem[\protect\citeauthoryear{Talon, Kumar \& Zahn}{Talon
  et~al.}{2002}]{tkz02}
Talon S.,  Kumar P.,    Zahn J.,  2002, ApJ, 574, 175

\bibitem[\protect\citeauthoryear{Thompson, Christensen-Dalsgaard, Miesch \&
  Toomre}{Thompson et~al.}{2003}]{th03}
Thompson M.,  Christensen-Dalsgaard J.,  Miesch M.,    Toomre J.,  2003, Annual
  Review of Astronomy and Astrophysics, 41, 599

\bibitem[\protect\citeauthoryear{Thompson, Toomre, Anderson, Antia, Berthomieu,
  Burtonclay, Chitre, J.Christensen-Dalsgaard, Corbard, Derosa
  et~al.,}{Thompson et~al.}{1996}]{th96}
Thompson M.,  Toomre J.,  Anderson E.,  Antia H.,  Berthomieu G.,  Burtonclay
  D.,  Chitre S.,  J.Christensen-Dalsgaard Corbard T.,  Derosa M.,    others .,
   1996, Science, 272, 1300

\bibitem[\protect\citeauthoryear{Zahn, Talon \& Matias}{Zahn
  et~al.}{1997}]{ztm97}
Zahn J.,  Talon S.,    Matias J.,  1997, A\&A, 322, 320

\end{thebibliography}
\section*{Acknowledgments}

We thank Gary Glatzmaier, Oliver Buhler and Jim McElwaine for helpful discussions.  Support for this research was provided by a NASA grant NNG06GD44G.  T.R. is supported by an NSF ATM Faculty Position in Solar Physics grant under award number 0457631 .  Computing resources were provided by NAS at NASA Ames. 

\clearpage
\begin{figure}
\centering
\includegraphics[width=6in]{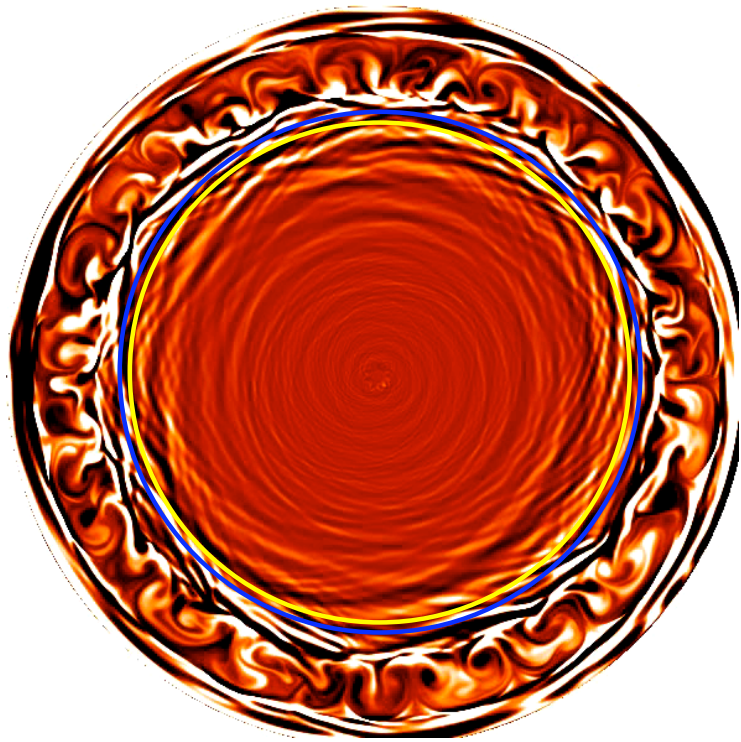}
\caption{Time snapshot of the temperature perturbation, with black representing cold material and white reprsenting hot material, with the initial field depicted by the overlaid lines.  Blue and yellow lines represent oppositely directed toroidal field.}
\end{figure}

\clearpage
\begin{figure}
\centering
\includegraphics[width=6in]{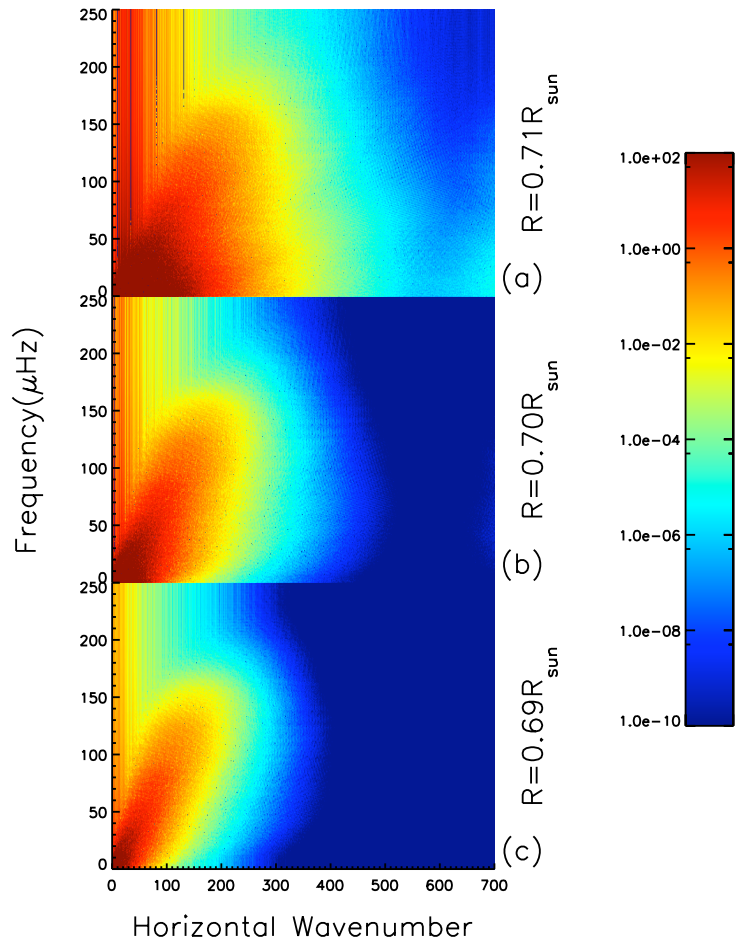}
\caption{Kinetic Energy Density spectrum for Model NF5 at 3 radii. The propagating waves seen in these models have lower frequencies than typical standing wave modes, which are short-lived and not seen here.  Over a very short depth, high wavenumber/low frequency waves are efficiently dissipated.}
\end{figure}

\clearpage
\begin{figure}
\centering
\includegraphics[width=6in]{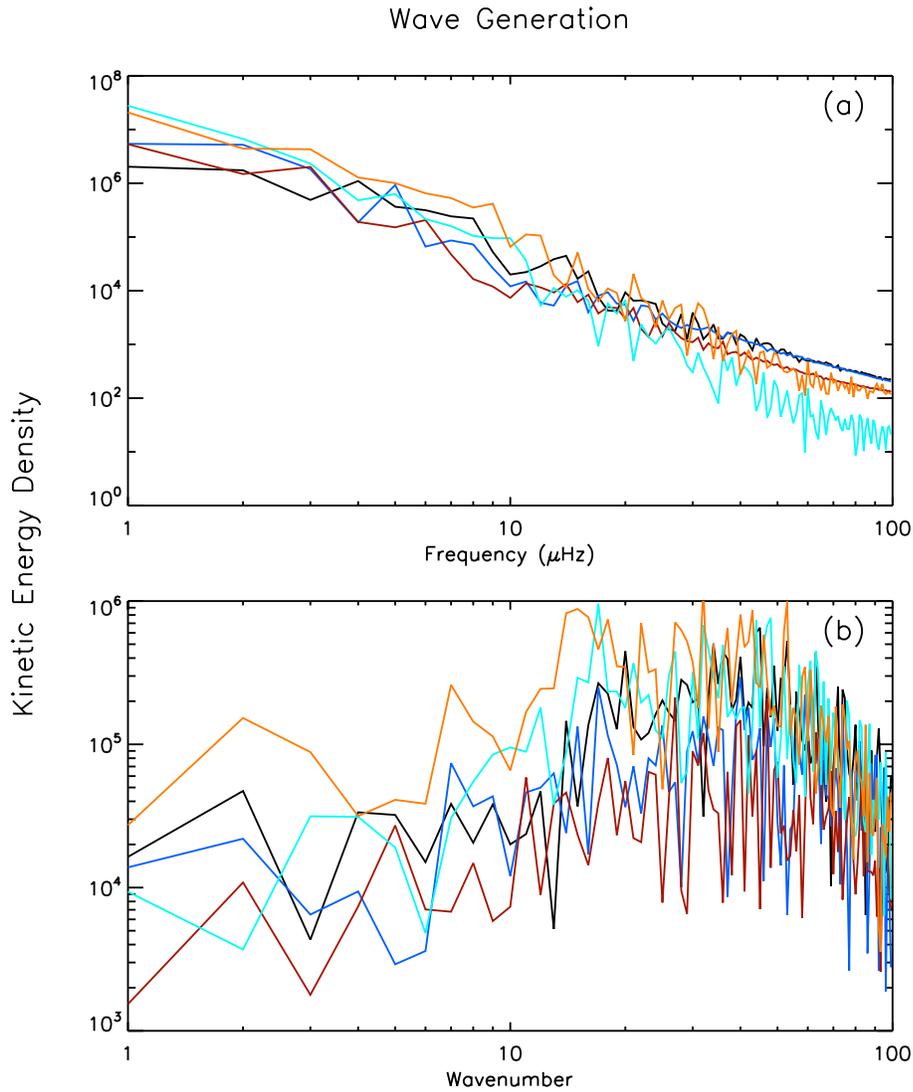}
\caption{Wave Generation.  This figure shows the kinetic energy density, calculated at 0.70R$_{\odot}$ as a function of frequency for wavenumber 10 (a) and wavenumber for a 10$\mu$Hz wave (b).  Different colors represent different models: orange is model KV2, cyan is KV1, black is NF5, blue is NF4 and red is NF2.  This figure shows the relative independence of wave generation on scale (b) and the strong dependence of that generation on frequency (a).}
\end{figure}
\clearpage
\begin{figure}
\centering
\includegraphics[width=6in]{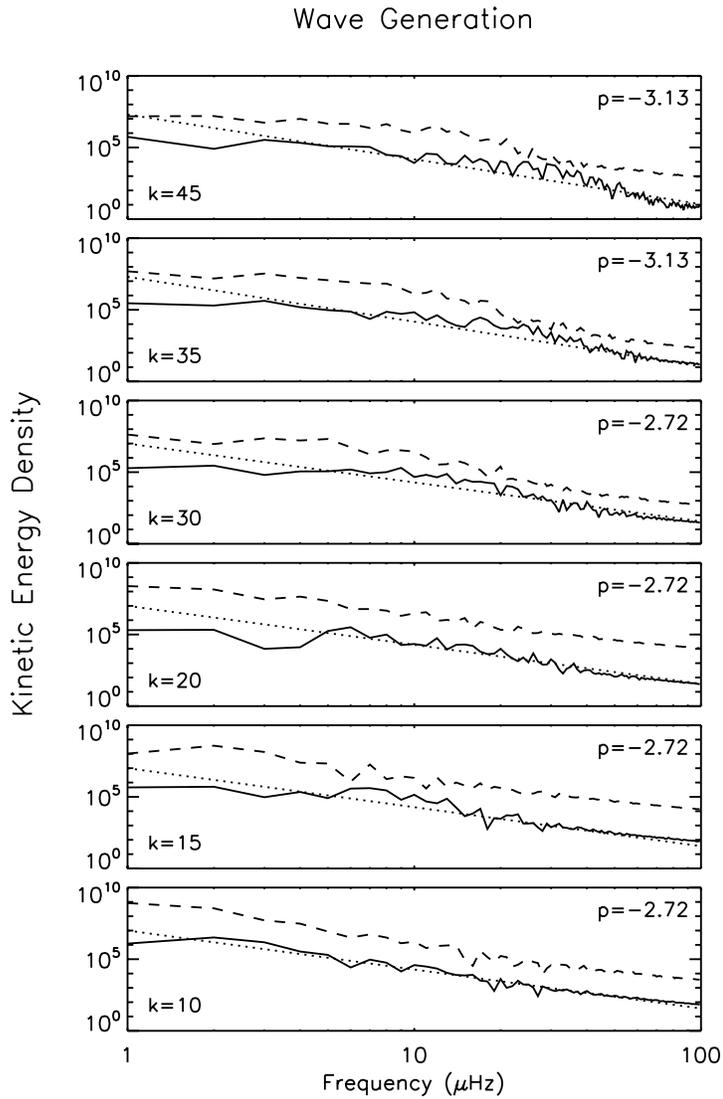}
\caption{Wave Generation.  Shown are wave kinetic energy density as a function of frequency for various wavenumbers, k at a radius of 0.69 R$_{\odot}$ (solid line, in the Radiation zone) and a radius of 0.80 R$_{\odot}$ (dashed line, in the Convection zone).  The frequency dependence can be fit well by a power law with exponent around -2.7 to -3.0 (dotted line).  Wave energy follows the distribution of convective energy quite well with very little difference except in amplitude.}
\end{figure}
\clearpage
\begin{figure}
\centering
\includegraphics[width=6in]{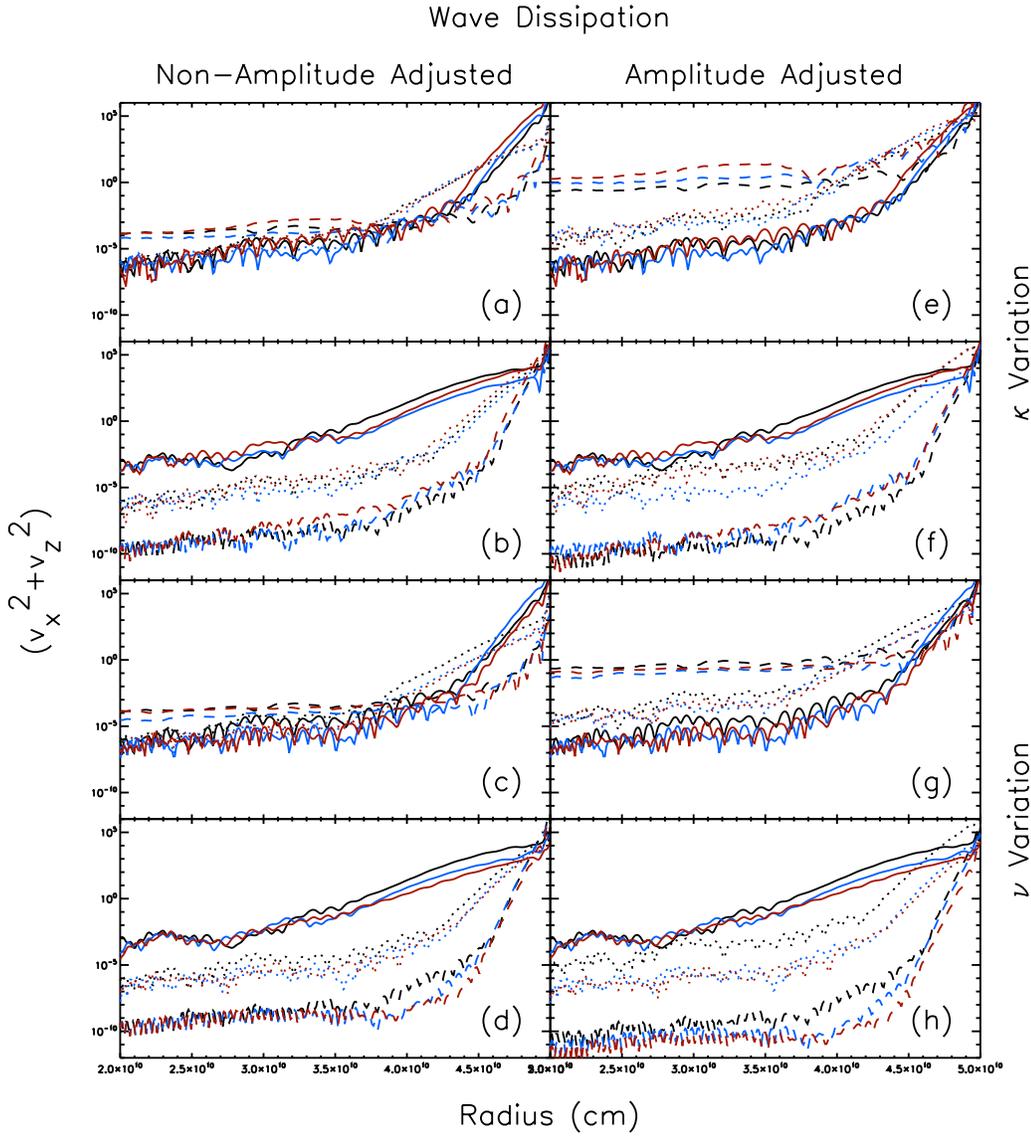}
\caption{Wave Dissipation.  Left hand panels show wave kinetic energy as a function of radius, with amplitudes at the base of the convection zone as calculated, while the right hand panels show the wave kinetic energy as a function of radius, with amplitudes at the base of the convection zone adjusted to be equivalent.  (a), (b), (e) and (f) show wave dissipation as a function of thermal diffusivity, $\kappa$, with black lines representing model NF5, blue representing KV1 and red representing KV2.  Different linetypes in (a) represent different frequencies with solid lines represent 5$\mu$Hz, dotted representing 20$\mu$Hz and dashed representing 80$\mu$Hz, all with horizontal wavenumber 10.  Different linetypes in (b) represent different wavenumbers, with solid lines representing k=3, dotted k=10 and dashed k=40, all with frequency of 10$\mu$Hz.  (c), (d), (g) and (h) show the same frequency/wavenumber combinations shown in (a), (b), (e) and (f) respectively, but different colors (c), (d), (g) and (h) representing models with varying viscous diffusivity, $\nu$, NF5 (black), NF4 (blue) and NF2 (red).  This figure shows that wave dissipation is a function of frequency, but is almost exactly countered by wave generation also being a function of frequency, leading to waves of different frequencies having very similar amplitudes in the deep radiative interior.}
\end{figure}
\clearpage
\begin{figure}
\centering
\includegraphics[width=6in]{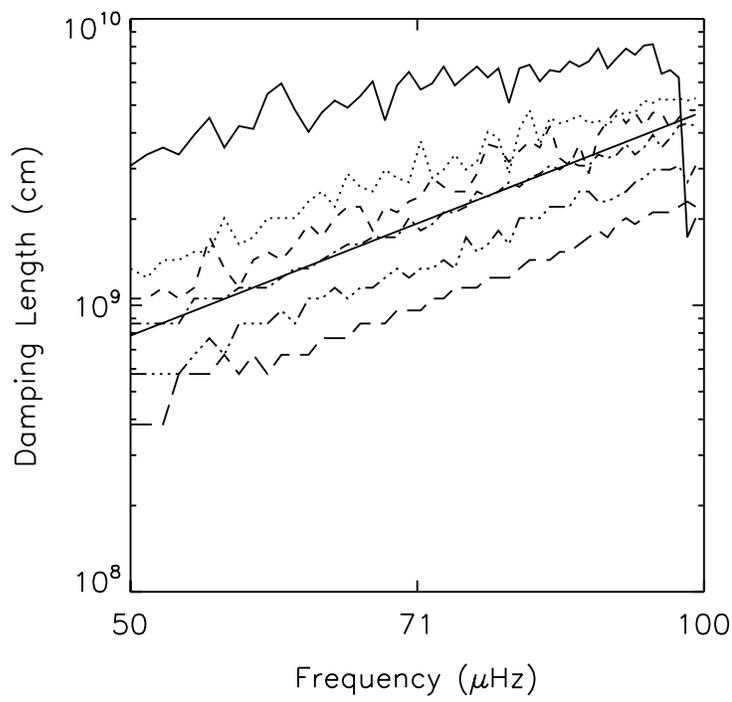}
\caption{Wave dissipation length.  Damping length measured in the simulations as the radius at which the wave amplitude dropped by a factor of ${\it e}$, for wavenumbers 15 (solid),18 (dotted), 22 (dashed), 23 (dash-dot), 25 (dash-triple dot) and 28 (long dash).  The straight line represents a power law of $\omega^{2.7}$.}
\end{figure}

\clearpage
\begin{figure}
\centering
\includegraphics[width=6in]{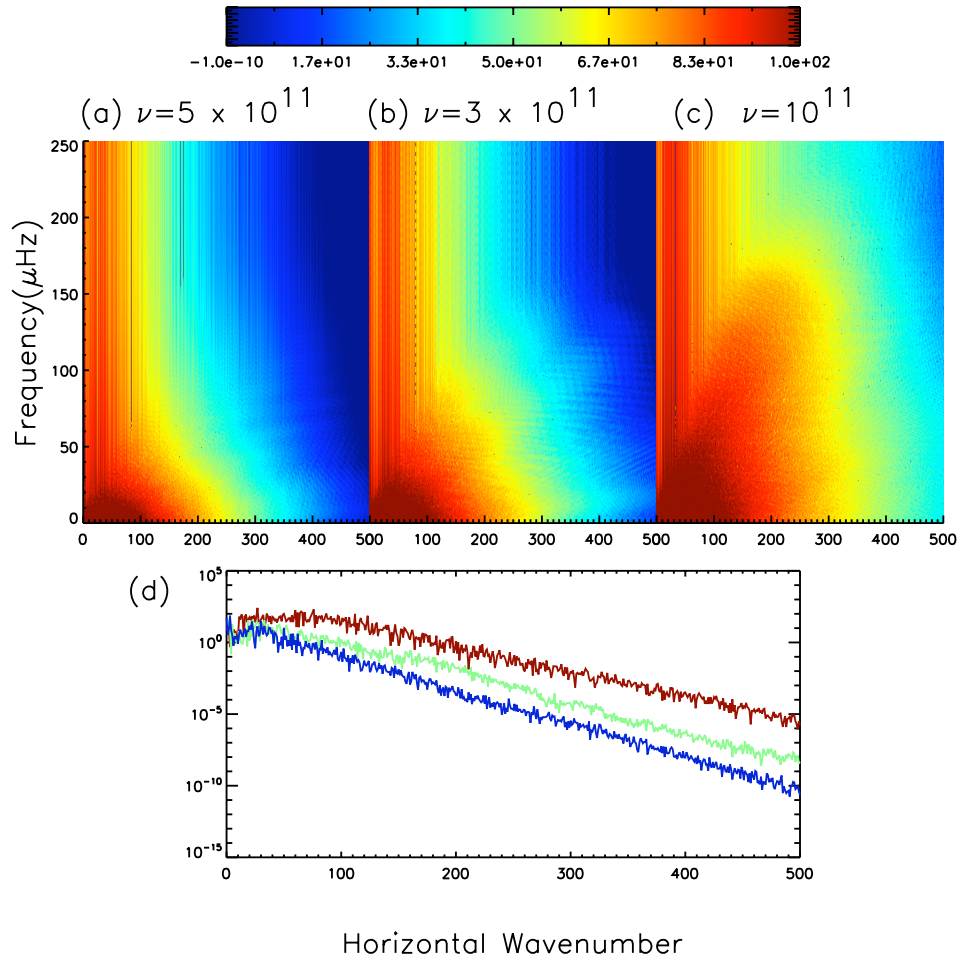}
\caption{Kinetic energy density spectrum for models NF1 (a), NF3 (b) and NF5 (c) at 0.70R$_{\odot}$, showing that lower values of $\nu$ lead to the generation of higher frequency waves.  (d) shows the amplitude of kinetic energy density as a function of horizontal wavenumber, for f=50$\mu$Hz, with the red line indicating model NF5, green line representing model NF3 and the blue line representing model NF1.}
\end{figure}

\clearpage
\begin{figure}
\centering
\includegraphics[width=6in]{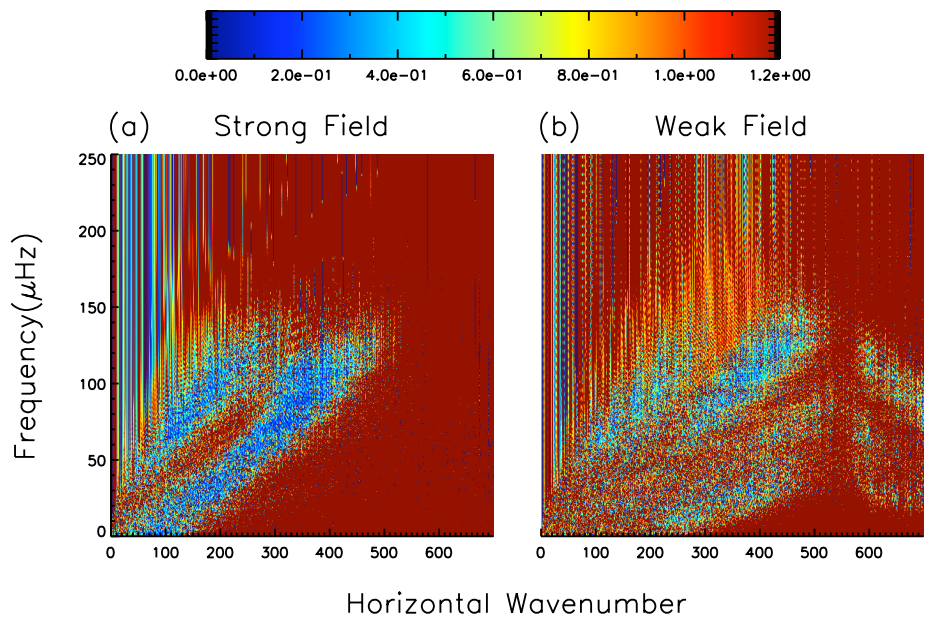}
\caption{The ratio of kinetic energy density for models BV1/NF1 (a) and for models BV2/NF1 in (b) at a radius of 0.69$R_{\odot}$.  Dark blue represents a ratio of zero, indicating there is substantially less kinetic energy in that region of (f,k) space in the magnetic case than in the non-magnetic case, while red represents a ratio of 1.2 indicating more energy in that region of (f,k) space in the magnetic model compared to the non-magnetic model.  This clearly demonstrates field strength dependent wave filtering.}
\end{figure}

\clearpage
\begin{figure}
\centering
\includegraphics[width=6in]{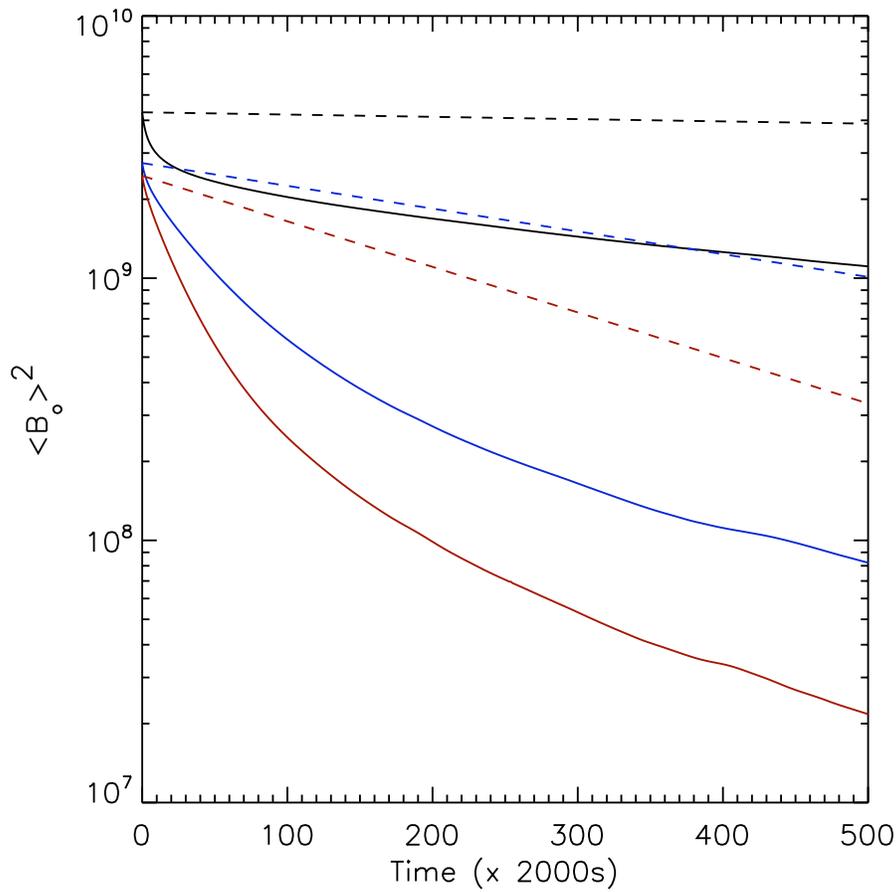}
\caption{The energy in the mean field as a function of time for models with varying values of the magnetic diffusivity $\eta$, EV1 (black), BV1 (blue) and EV2 (red).  Dashed lines represent the diffusion expected for dissipation of the large scale field calculated assuming $\tau_{diff}=D^{2}/\eta$, where D is the initial depth of the magnetic layer.}
\end{figure}

\clearpage
\begin{figure}
\includegraphics[width=6in]{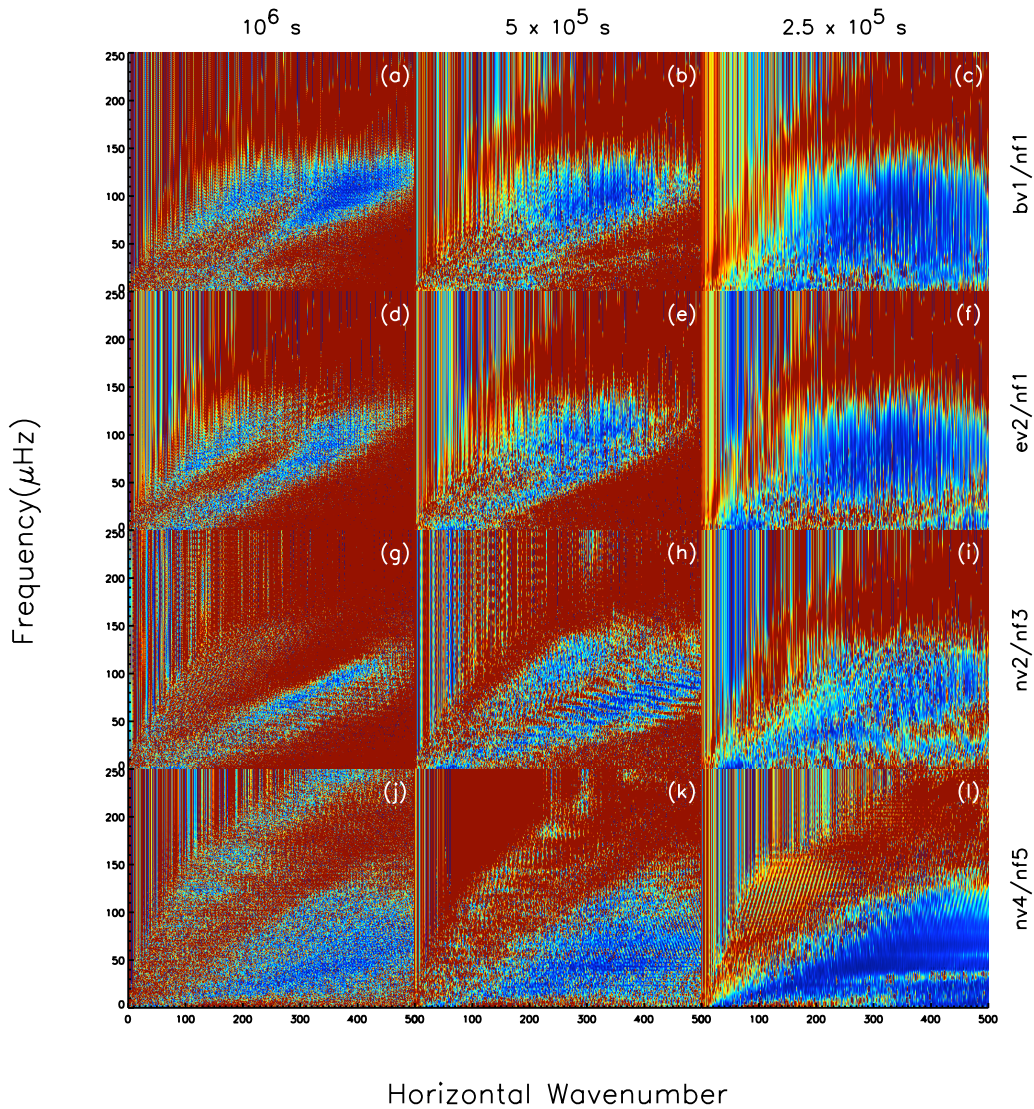}
\caption{Wave filtering in the presence of field.  This image shows kinetic energy density ratio spectra for different models over different times.  Different rows represent filtering for different models as indicated in the figure.  Different columns represent the filtered spectrum over different time intervals with the left-most plots (a,d,g,j) representing integrations over 10$^{6}$s, (b,e,h,k) representing integration over $5\times 10^{5}$s and (c,f,i,l) representing integrations over only $2.5\times 10^{5}$s.  One can see that at early times, when the field has not decayed, the filtering is strong in all cases.  However, after the field has decayed more (left most plots) the effect of filtering has decreased.  Note that all of these models have the same (strong) initial field strengths, but varying degrees of dissipation by both magnetic and viscous diffusivity.}
\end{figure}

\clearpage
\begin{figure}
\centering
\includegraphics[width=6in]{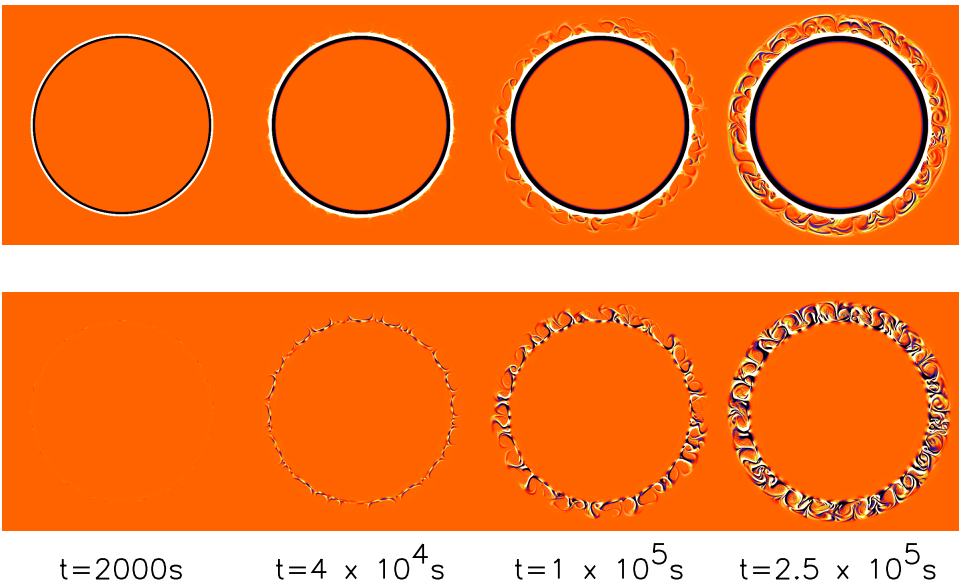}
\caption{Time snapshots of the magnetic field evolution.  The top panel shows the toroidal field, while the bottom panel shows the radial field, both progressing in time to the right.  The first snapshot of the radial field shows no structure as the field initially imposed is purely toroidal.  Radial field is created in time as overshooting plumes distort and entrain the field.}
\end{figure}

\clearpage
\begin{figure}
\centering
\includegraphics[width=6in]{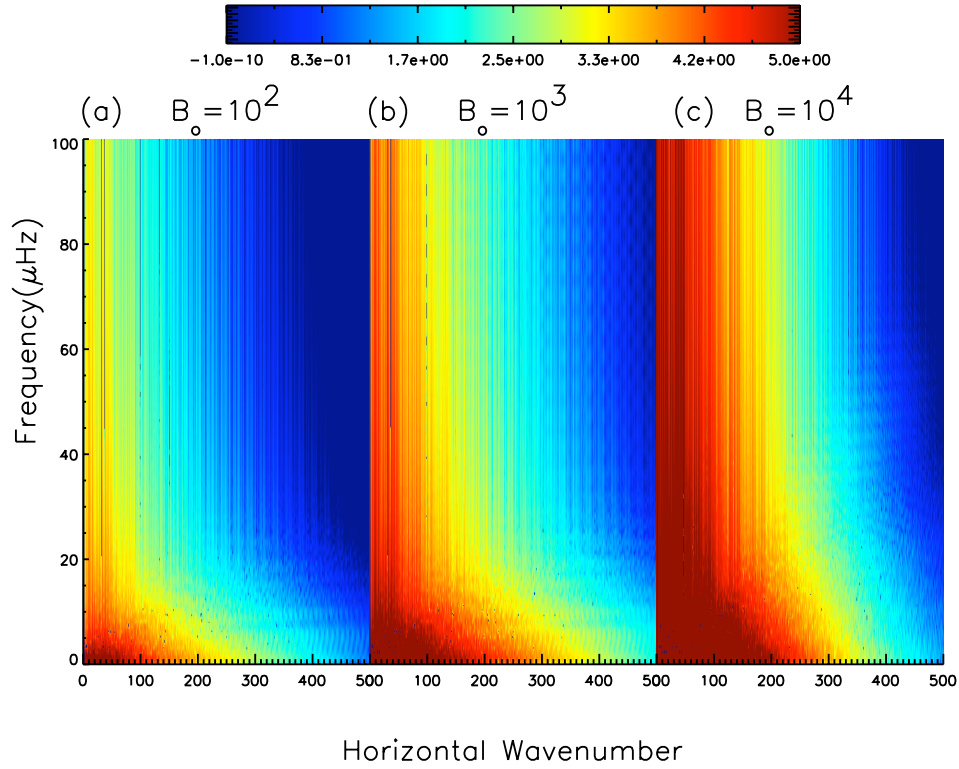}
\caption{Magnetic Energy spectrum at 0.70R$_{\odot}$ as a function of field strength, with (a) showing the lowest field strength and (c) showing the highest field strength.  As expected from the simple dispersion relation for Alfven waves the highest field strength supports the highest frequency waves.}
\end{figure}

\clearpage
\begin{figure}
\includegraphics[width=6in]{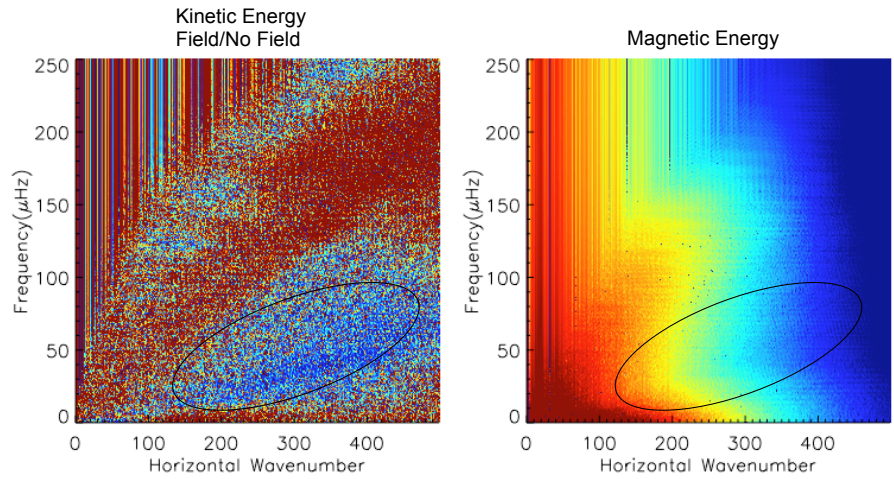}
\caption{Wave filtering in the presence of field.  On the left we show the ratio of kinetic energy in the field case to the non-field case (same as what is shown in figure 8), on the right we show the magnetic energy spectrum.  The circled areas show regions where there is little kinetic energy in the magnetic cases and similarly, on the right, where there is little magnetic energy in the magnetic spectrum.  In this model the magnetic Prandtl number is less than one and the field is sufficiently tied to the flow so that the region of (f,k) space where there is little kinetic energy is also that region of (f,k) space where there is little magnetic energy.}
\end{figure}

\clearpage
\begin{figure}
\centering

\includegraphics[width=6in]{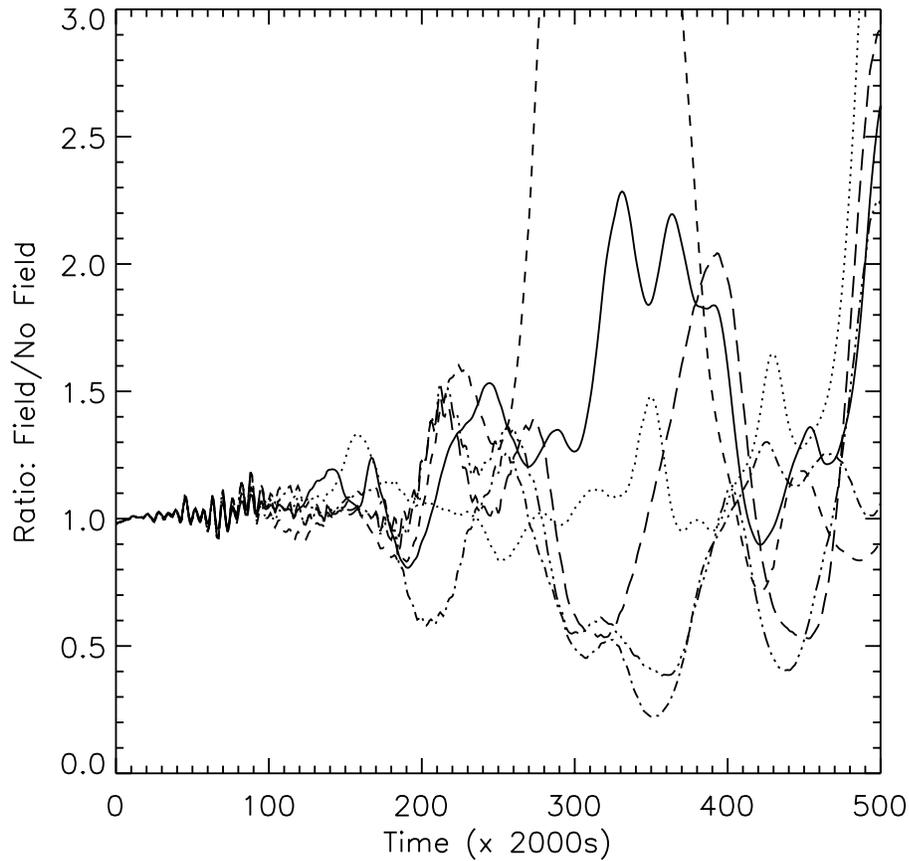}
\caption{Integrated wave energy in the deep radiative interior.  In this figure we show the ratio of kinetic energy density for the magnetic case to the non-magnetic case integrated from the base of the computational domain to 0.67R$_{\odot}$.  Different linetypes represent various models: solid line NV4/NF5, dashed line NV3/NF4, dotted line NV2/NF3, dash-dotted line NV1/NF2, dash-three dot line BV1/NF2 and long dashed line BV2/NF2.  Initially the models all oscillate around one, but over time the models diverge with substantial excursions ranging from substantially less energy in the interior in the field case to substantially more.(b) The same ratio shown in (a) NV4/NF5 (solid line) and the sum of kinetic and magnetic wave energy ratio for NV4/NF5 in the field region, multipled by two for ease of comparison.}
\end{figure}

\clearpage
\begin{figure}
\includegraphics[width=6in]{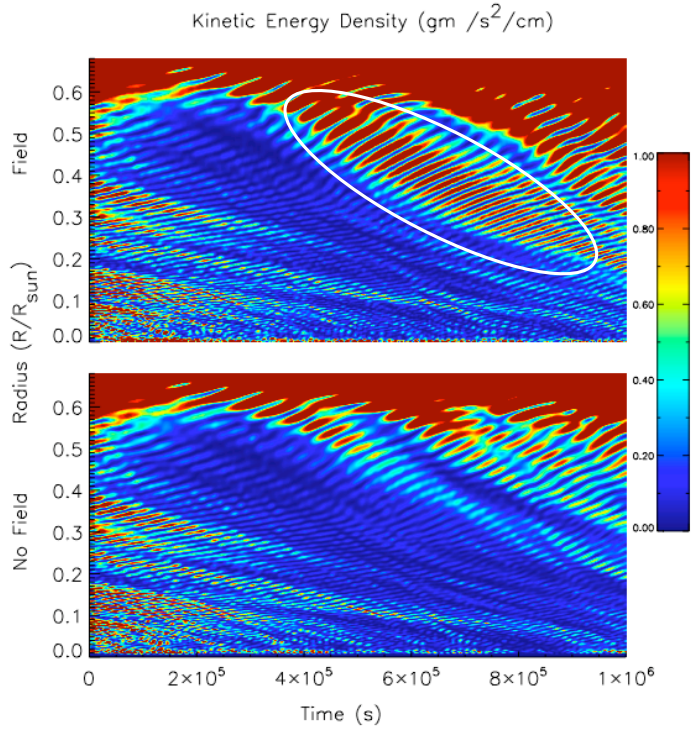}
\caption{IGW generation by magnetic field.  This figure shows the integrated wave energy as a function of time (x-axis) and radius (y-axis) beneath the magnetic field.  On the bottom we show the unmagnetized Model NF5, while on the top we show the magnetic Model NV4.  Initially (to about 2$\times 10^{5}s$) the two models are identical, subsequently one can see that the magnetic model produces more kinetic energy in the deep interior.  The circled region shows increased kinetic energy in a group of IGW.  Subsequent motion in the magnetized model also has larger kinetic energy.}
\end{figure}

\clearpage
\begin{figure}
\includegraphics[width=6in]{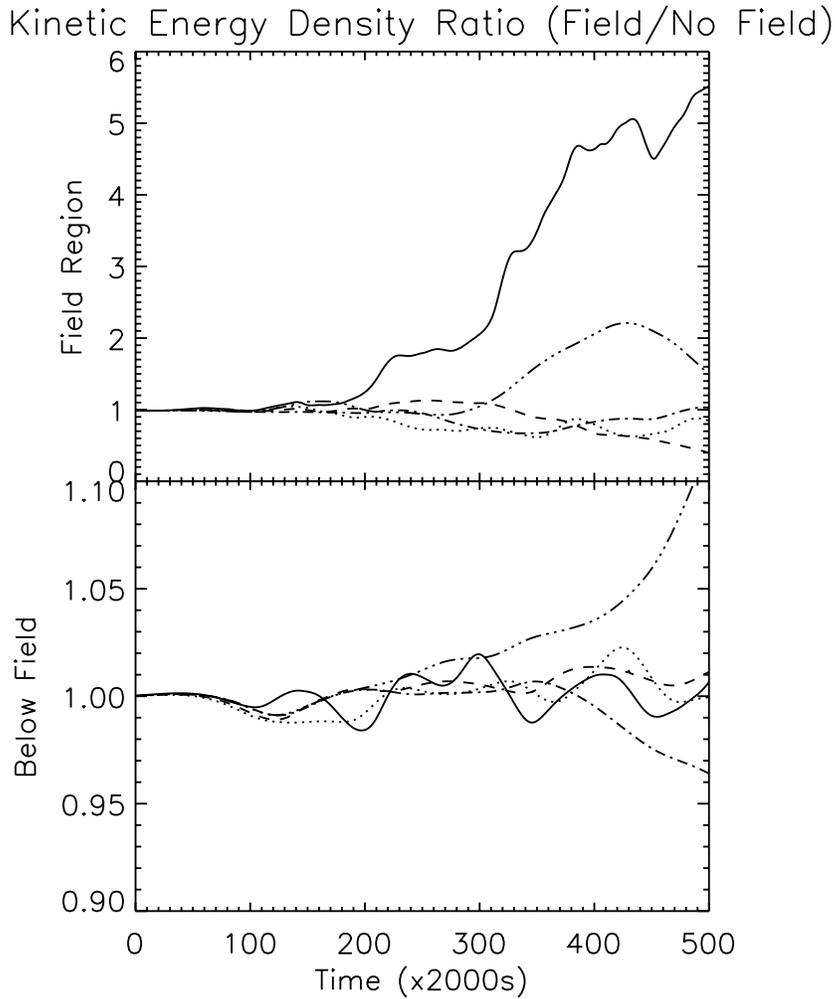}
\caption{Integrated mean flow in the radiative interior.  In this figure we show the ratio of the kinetic energy in mean flow for the magnetized case to the unmagnetized case for various models, solid line represents NV4/NF5, dotted line is NV3/NF4, dashed line is NV2/NF3, dash- dot line is BV1/NF1 and dash-three dot line is EV1/NF1.  In the top panel we show this ratio in the field region (integrated from 0.67R$_{\odot}$ to 0.70R$_{\odot}$), and in the bottom panel we show this ratio below the field (integrated from the bottom of the domain to 0.67R$_{\odot}$).  Below the field we see that this ratio is nearly one, with some minor excursions above and below one.  However, in the field region we see that both the model with the lowest $\nu$ (NV4/NF5) and that with the lowest $\eta$ (EV1/NF1) show ratios above one.}
\end{figure}

\end{document}